\documentclass[12pt]{article}
\usepackage{arydshln}
\usepackage{amssymb,graphicx}
\usepackage{amsmath}
\title{A Tutorial Introduction to the Lambda Calculus}
\author{ Raul  Rojas\thanks{Send corrections or suggestions to rojas@inf.fu-berlin.de}}
\date{\mbox{Freie Universit\"at Berlin}\\
Version 2.0, 2015}
\begin{document}
\maketitle

\begin{abstract}
This paper is a concise and painless introduction to the $\lambda$-calculus. This formalism  
 was developed by Alonzo Church as a tool for studying the mathematical properties of effectively computable functions. The formalism 
became popular and has provided  a strong theoretical foundation  for the family of functional programming languages. 
This tutorial shows how to perform arithmetical and logical computations using the  $\lambda$-calculus and how to define recursive functions, 
even though  $\lambda$-calculus functions are unnamed and thus cannot refer explicitly to themselves. 
\end{abstract}

\section{Definition}

The $\lambda$-calculus can be called the {\it smallest universal programming language in the world\/}. The $\lambda$-calculus consists of a single transformation rule (variable substitution, also called $\beta$-conversion) and a single function definition scheme. It was introduced in the 1930s by Alonzo Church as a way of formalizing the concept of effective computability. The $\lambda$-calculus is universal in the sense that any computable function can be expressed and evaluated using this formalism. It is thus equivalent to Turing machines. However, the $\lambda$-calculus emphasizes the use of symbolic transformation rules and does not care about the actual machine implementation. It is an approach more related to software than to hardware. 

The central concept in $\lambda$-calculus is that of ``expression''. 
A ``name'' is an identifier which, for our purposes, can be any of the letters $a,b,c$, etc. 
An expression can be just a name or can be a function. Functions use the Greek letter $\lambda$ to mark the
name of the function's arguments. The ``body'' of the function specifies how the arguments are to be rearranged. The identity function, for example, is represented by the string  $(\lambda x.x)$. The fragment ``$\lambda x$'' tell us that the function's argument is $x$, which is returned unchanged as ``$x$'' by the function.

Functions can be applied to other functions. The function {\sf A}, for example, applied to the function {\sf B}, would be written as {\sf AB}. In this tutorial, capital letters are used to represent functions. In fact, anything of interest in $\lambda$-calculus is a function. Even numbers or logical values will be represented by functions that can act on one another in order to transform a string of symbols into another string. There are no types in $\lambda$-calculus: any function can act on any other. The programmer is responsible for keeping the computations sensible.

An expression is defined recursively as follows:
\begin{eqnarray*}
<{\mbox{expression}}> &:=& <{\mbox{name}}> \mid <{\mbox{function}}> \mid <{\mbox{application}}>\\
<{\mbox{function}}> &:=& \lambda <{\mbox{name}}>. <{\mbox{expression}}>\\
<{\mbox{application}}> &:=& <{\mbox{expression}}><{\mbox{expression}}>
\end{eqnarray*}

An expression can be surrounded by parenthesis for clarity, that is, if {\sf E} is an expression, ({\sf E}) is the same expression. Otherwise, the only keywords used in the language are $\lambda$ and the dot. In order to avoid cluttering expressions with parenthesis, we adopt the convention that function application associates from the left, that is, the composite expression
$${\sf E}_1{\sf E}_2{\sf E}_3\ldots {\sf E}_n$$
is evaluated applying the successive expressions as follows
$$\biggl(\ldots\bigl(({\sf E}_1{\sf E}_2){\sf E}_3\bigr)\ldots {\sf E}_n\biggr)$$
As can be seen from the definition of $\lambda$-expressions, a well-formed example of a function is the previously mentioned string, enclosed or not in parentheses:
$$
\lambda x.x \equiv (\lambda x.x)
$$
We use the equivalence symbol ``$\equiv$'' to indicate that when ${\sf A}\equiv {\sf B}$, ${\sf A}$ is just a synonym for  ${\sf B}$. As explained above, the name right after the $\lambda$ is the identifier of the argument of this function. The expression after the point (in this case a single $x$) is called the ``body'' of the function's definition.

\begin{figure}[htb]
\centerline{\includegraphics[width=4cm]{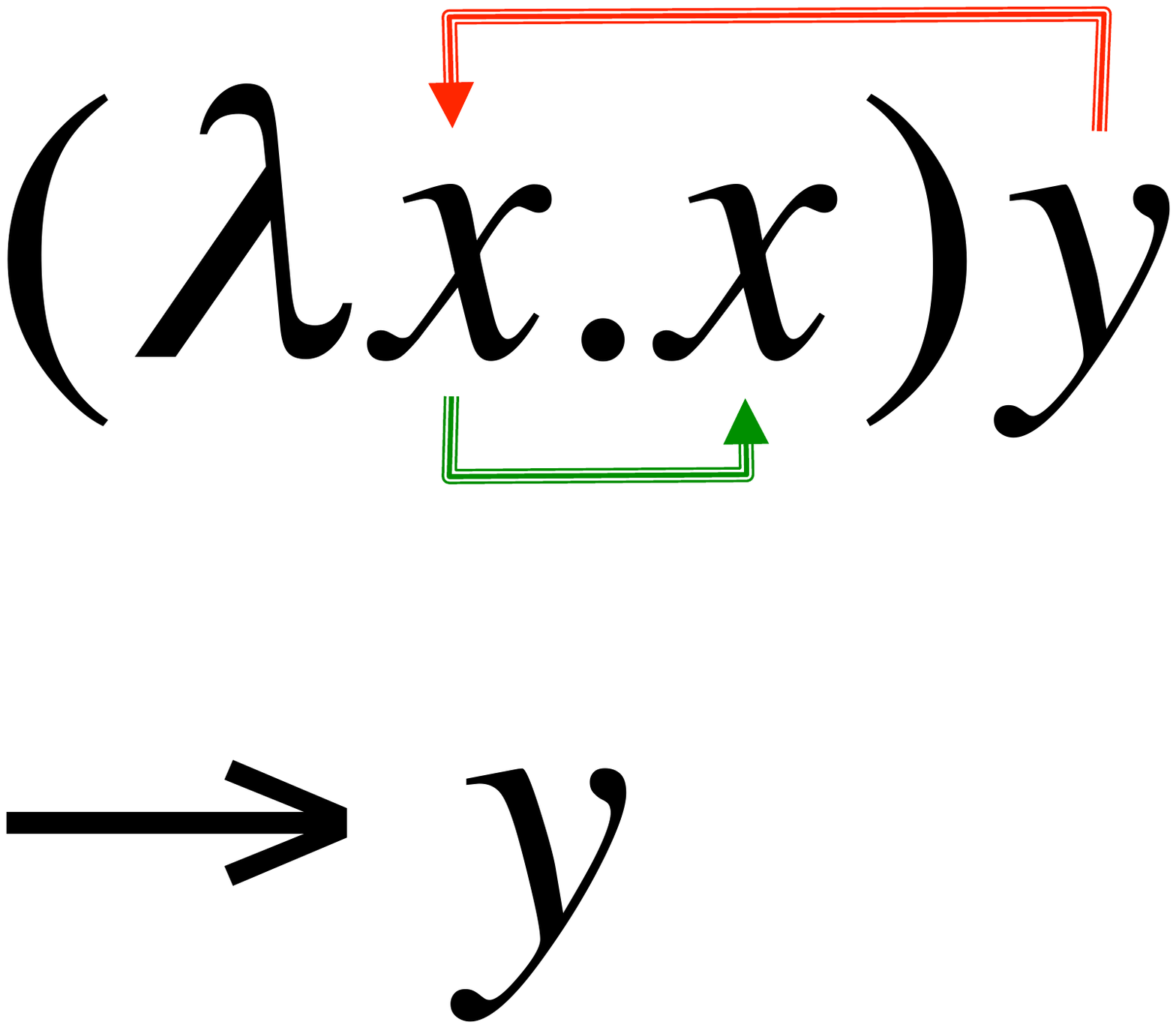}\includegraphics[width=4cm]{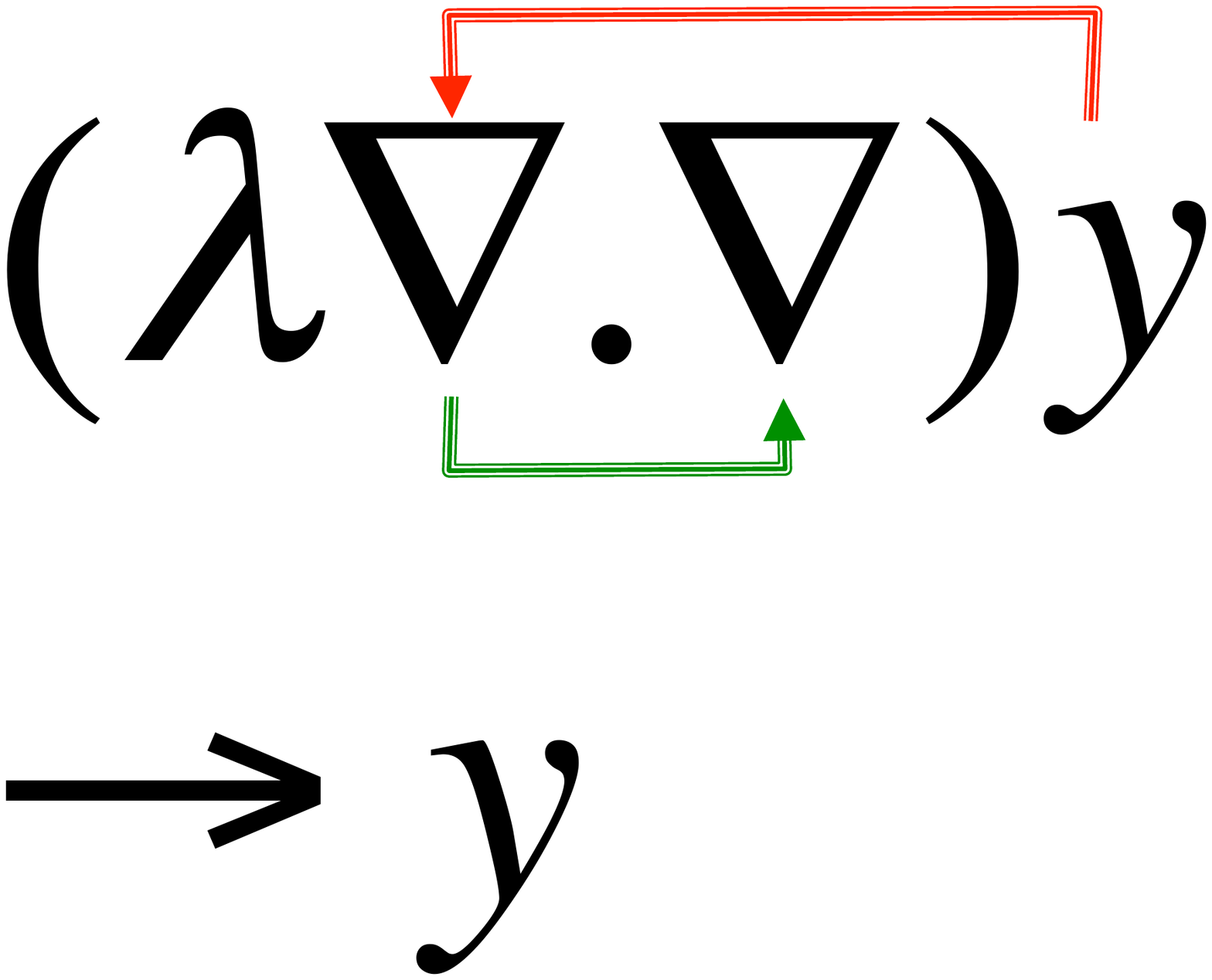}}
\caption{The same reduction shown twice. The symbol for the function's argument is just a place holder.\label{fig1}}
\end{figure}

Functions can be applied to expressions. A simple example of an application is 
$$(\lambda x.x)y$$
This is the identity function applied to the variable $y$. Parenthesis help to avoid ambiguity. Function applications are evaluated by substituting the ``value'' of the argument $x$ (in this case the $y$ being processed) in the body of the function definition. Fig.~\ref{fig1} shows how the variable $y$ is ``absorbed'' by the function (red line), and also shows where it is used as a replacement for $x$ (green line). The result is a reduction, represented by the right arrow, with the final result $y$.

Since we cannot always have pictures, as in Fig.~\ref{fig1}, the notation $[y/x]$ is used to indicate that all occurrences of $x$ are substituted by $y$ in the function's body. We write, for example, $(\lambda x.x)y \rightarrow [y/x]x \rightarrow y$. The names of the arguments in function definitions do not carry any meaning by themselves. They are just ``place holders'', that is, they are used to indicate how to rearrange the arguments of the function when it is evaluated. Therefore all the strings below represent the same function:
$$(\lambda z.z) \equiv (\lambda y.y)\equiv (\lambda t.t)\equiv (\lambda u.u)
$$
This kind of purely alphabetical substitution is also called $\alpha$-reduction.

\subsection{Free and bound variables}

If we only had pictures of the plumbing of $\lambda$-expressions, we would not have to care about the names of variables. Since we are using letters as symbols, we have to be careful if we repeat them, as shown in this section.

\begin{figure}[htb]
\centerline{\includegraphics[width=7cm]{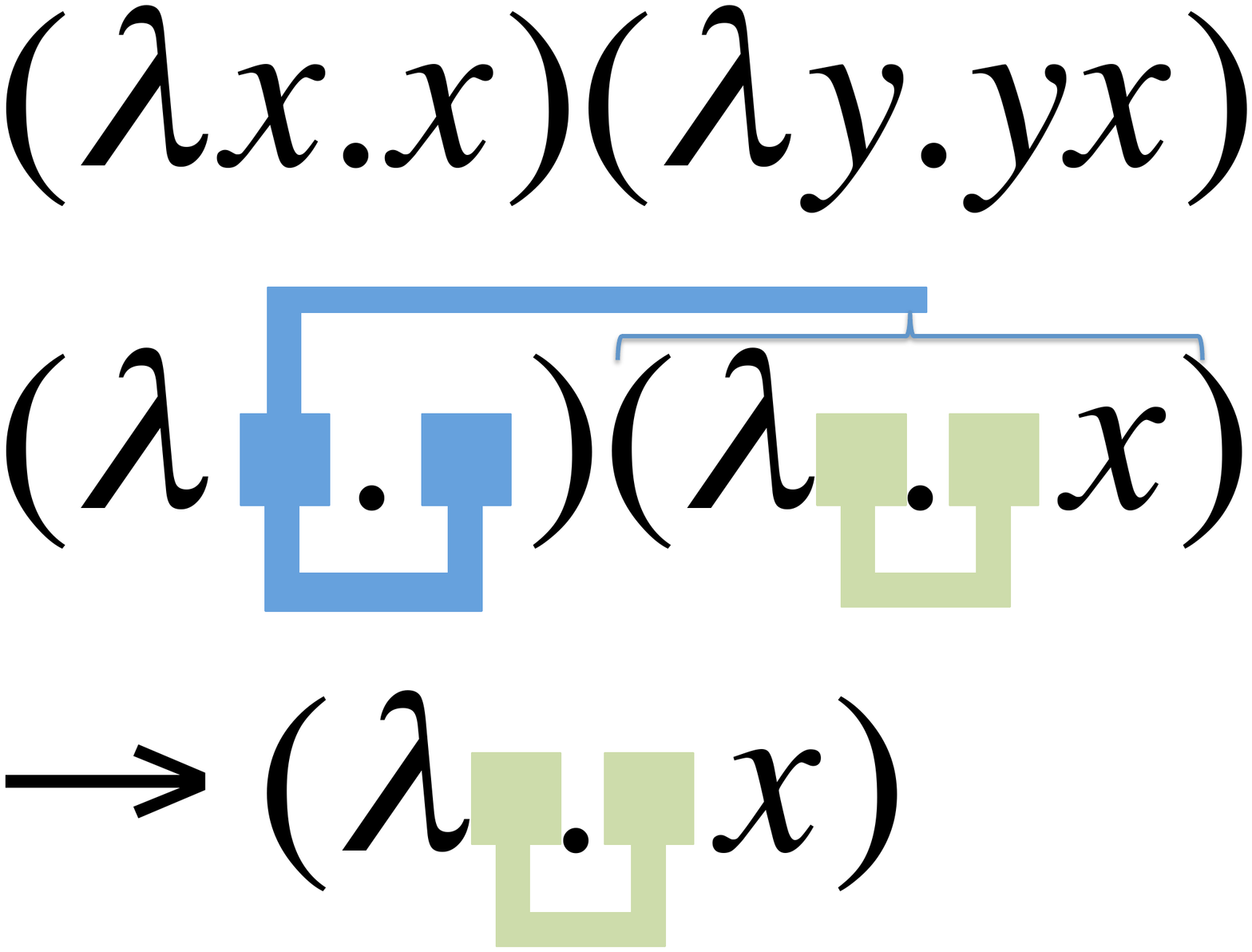}}
\caption{In successive rows: The function application, the ``plumbing'' of the symbolic expression, and the resulting reduction.\label{fig2}}
\end{figure}

In $\lambda$-calculus all names are local to definitions (like in most programming languages). In the function $\lambda x.x$ we say that $x$ is ``bound'' since its occurrence in the body of the definition is preceded by $\lambda x$. A name not preceded by a $\lambda$ is called a ``free variable''. In the expression 
$$(\lambda {\bf x}.{\bf x})(\lambda{\bf y}.{\bf y}x)$$ 
the $x$ in the body of the first expression from the left is bound to the first $\lambda$. The $y$ in the body of the second expression is bound to the second $\lambda$, and the following $x$ is free. Bound variables are shown in bold face.  It is very important to notice that this $x$ in the second expression is totally independent of the $x$ in the first expression. This can be more easily seen if we draw the ``plumbing'' of the function application and the consequent reduction, as shown in Fig.~\ref{fig2}.

In Fig.~\ref{fig2} we see how the symbolic expression (first row) can be interpreted as a kind of circuit, where the bound argument is moved to a new position inside the body of the function. The first function (the identity function) ``consumes'' the second one. The symbol $x$ in the second function has no connections with the rest of the expression, it is floating free inside the function definition.

Formally, we say that a variable $<$name$>$ is free in an expression if one of the following three cases holds:
\begin{itemize}
\item[$\bullet$] $<$name$>$ is free in $<$name$>$. \\(Example: $a$ is free in $a$).
\item[$\bullet$] $<$name$>$ is free in $\lambda$$<$name$_1>$. $<$exp$>$ if the identifier $<$name$>\neq$$<$name$_1>$ and $<$name$>$ is free in $<$exp$>$. \\(Example: $y$ is free in $\lambda x.y$).
\item[$\bullet$] $<$name$>$ is free in $E_1E_2$ if $<$name$>$ is free in  $E_1$ or if it is free in $E_2$.\\ (Example: $x$ is free in $(\lambda x.x)x$).
\end{itemize}

 A variable $<$name$>$ is bound if one of two cases holds:
\begin{itemize}
\item[$\bullet$]  $<$name$>$ is bound in $\lambda$ $<$name$_1>$. $<$exp$>$ if the identifier $<$name$>=$$<$name$_1>$ or if $<$name$>$ is bound in  $<$exp$>$.\\
(Example: $x$ is bound in $(\lambda y.(\lambda x.x))$).
\item[$\bullet$] $<$name$>$ is bound in $E_1E_2$ if $<$name$>$ is bound in $E_1$ or if it is bound in $E_2$.\\
(Example: $x$ is bound in $(\lambda x.x)x)$.
\end{itemize}
It should be emphasized that the same identifier can occur free and bound in the same expression. In the expression 
$$(\lambda {\bf x}.{\bf x}y)(\lambda {\bf y}.{\bf y})$$
 the first $y$ is free in the parenthesized subexpression to the left, but it is bound in the subexpression to the right. Therefore, it occurs free as well as bound in the whole expression (the bound variables are shown in bold face).
 
\subsection{Substitutions}

The more confusing part of standard $\lambda$-calculus, when first approaching it,is the fact that we do not give names to functions. Any time we want to apply a function, we just write the complete function's definition  and then proceed to evaluate it. To simplify the notation, however, we will use capital letters, digits and other symbols (san serif) as synonyms for some functions. The identity function, for example, can be denoted by the letter {\sf I}, using it as shorthand for $(\lambda x.x)$.

The identity function applied to itself is the application $${\sf I}{\sf I}\equiv (\lambda x.x)(\lambda x.x).$$ In this expression, the first $x$ in the body of the first function in parenthesis is independent of the $x$ in the body of the second function (remember that the ``plumbing'' is local). Just to emphasize the difference we can in fact rewrite the above expression as $${\sf I}{\sf I}\equiv (\lambda x.x)(\lambda z.z).$$  The identity function applied to itself $${\sf I}{\sf I}\equiv (\lambda x.x)(\lambda z.z)$$ yields therefore
$$[(\lambda z.z)/x]x\rightarrow\lambda z.z \equiv{\sf I},$$ that is, the identity function again.

\begin{figure}[htb]
\centerline{\includegraphics[width=10cm]{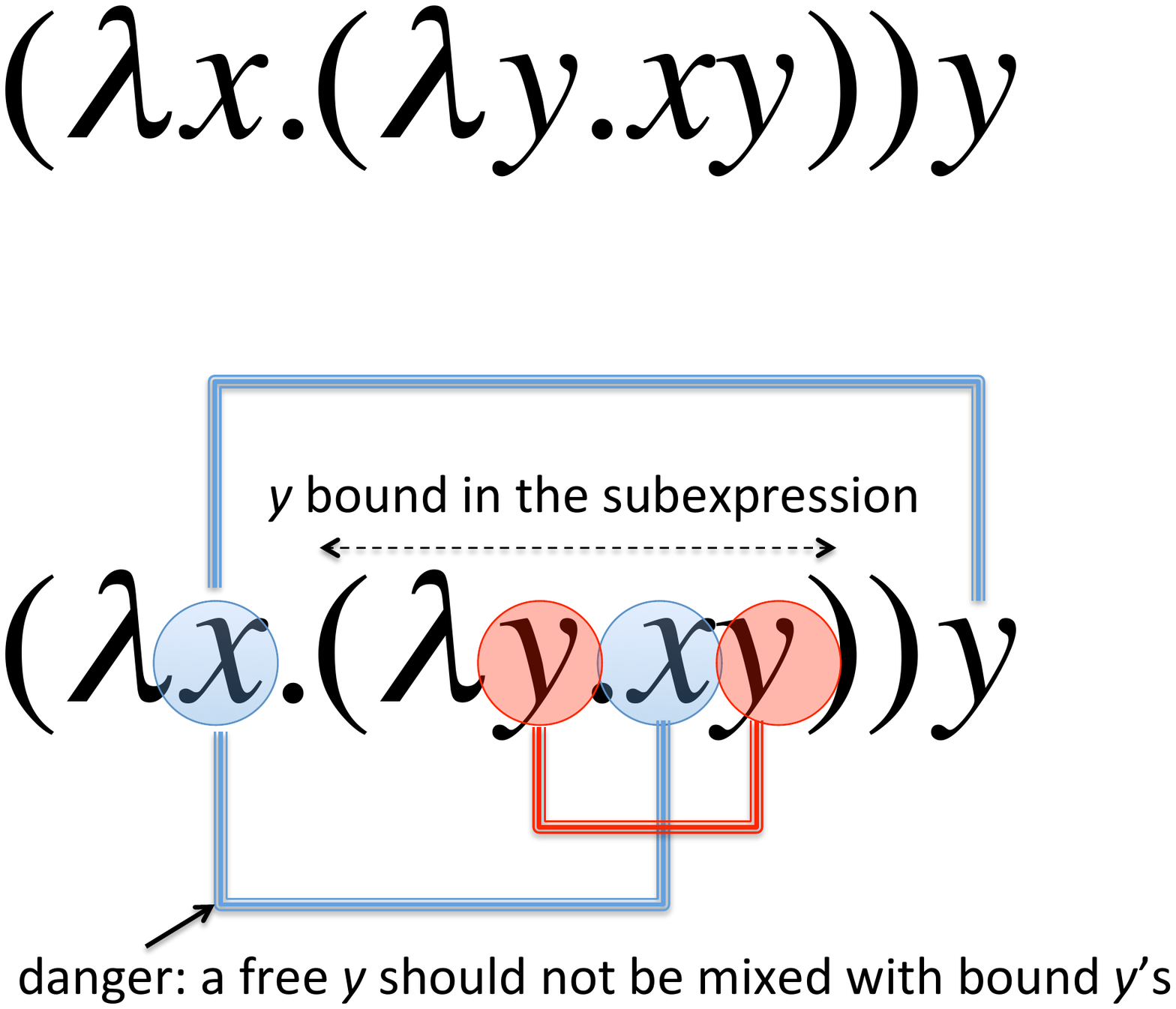}}
\caption{A free variable should not be substituted in a subexpression where it is bound, otherwise a new ``plumbing'', different to the original, would be generated.\label{fig4}}
\end{figure} 

When performing substitutions, we should be careful  to avoid mixing up free occurrences of an identifier with bound ones. In the expression $$\bigl(\lambda {\bf x}.(\lambda {\bf y}.{\bf x}{\bf y})\bigr)y$$ the function on the left contains a bound $y$, whereas the $y$ on the right is free. An incorrect substitution would mix the two identifiers in the erroneous result $$(\lambda {\bf y}.y{\bf y}).$$ Simply by renaming the bound $y$ to $t$ we obtain $$\bigl(\lambda x.(\lambda t.xt)\bigr)y \rightarrow (\lambda t.yt)$$
which is a completely different result but nevertheless the correct one. 

Therefore, if the function $\lambda x.<\mbox{exp}>$ is applied to $E$, we substitute all {\it free\/} occurrences of $x$ in $<\mbox{exp}>$ with $E$. If the substitution would bring a free variable of $E$ in an expression where this variable occurs bound, we rename the bound variable before performing the substitution. For example, in the expression
$$
\left(\lambda x.(\lambda y.(x(\lambda x.xy)))\right)y
$$
we associate the first $x$ with $y$. In the body
$$\left(\lambda y.(x(\lambda x.xy))\right)$$
only the first $x$ is free and can be substituted. Before substituting though, we have to rename the variable $y$ to avoid mixing its bound with its free occurrence:
$$
[y/x]\left(\lambda t.(x(\lambda x.xt))\right) \rightarrow \left(\lambda t(y(\lambda x.xt))\right)
$$
In normal order reduction we reduce always the left most expression of a series of applications first. We continue until no further reductions are possible.

\section{Arithmetic}
A programming language should be capable of specifying arithmetical calculations. Numbers can be represented in the $\lambda$-calculus starting from zero and writing ``successor of zero'', that is ``suc(zero)'', to represent 1, ``suc(suc(zero))'' to represent 2, and so on. Since in $\lambda$-calculus we can only define new functions, numbers will be defined as functions using the following approach: zero can be defined as $$\lambda s.(\lambda z.z)$$ This is a function of two arguments $s$ and $z$. We will abbreviate such expressions with more than one argument as $$\lambda sz.z$$ It is understood here that $s$ is the first argument to be substituted during the evaluation and $z$ the second. Using this notation, the first natural numbers can be defined as 
\begin{eqnarray*}
{\sf 0} &\equiv& \lambda sz.z\\
{\sf 1} &\equiv& \lambda sz.s(z)\\
{\sf 2} &\equiv & \lambda s z.s(s(z))\\
{\sf 3} &\equiv&  \lambda s z.s (s(s(z)))
\end{eqnarray*}
and so on.

The big advantage of defining numbers in this way is that we can now apply a function $f$ to an argument $a$ any number of times. For example, if we want to apply $f$ to $a$ three times we apply the function {\sf 3} to the arguments $f$ and $a$ yielding:
$$
{\sf 3}fa \rightarrow  (\lambda sz. s(s(sz)))fa \rightarrow f(f(fa)).
$$
This way of defining numbers provides us with a language construct similar to an instruction such as ``FOR i=1 to 3'' in other languages. The number zero applied to the arguments $f$ and $a$ yields ${\sf 0}fa \equiv (\lambda sz.z)fa \rightarrow a$. That is, applying the function $f$ to the argument $a$ {\em zero times} leaves the argument $a$ unchanged.

Our first interesting function, after having defined the natural numbers, is the successor function. This can be defined as 
$${\sf S}\equiv\lambda nab.a(nab).$$
The definition looks awkward but it works. For example, 
the successor function applied to our representation for zero is the expression:
$${\sf S0}\equiv (\lambda nab.a(nab)){\sf 0}$$ 
In the body of the first expression we substitute the occurrence of $n$ with ${\sf 0}$ and this produces the reduced expression:
$$\lambda ab.a({\sf 0}ab) \rightarrow \lambda ab.a(b)\equiv {\sf 1}$$
That is, the result is the representation of the number {\sf 1}  (remember that bound variable names are ``dummies'' and can be changed). 

Successor applied to {\sf 1} yields:
$${\sf S}{\sf 1}\equiv (\lambda nab.a(nab)){\sf 1} \rightarrow \lambda ab.a ({\sf 1}ab) \rightarrow \lambda ab.a(ab)\equiv {\sf 2}$$
Notice that the only purpose of applying the number ${\sf 1}\equiv(\lambda sz.sz)$ to the arguments $a$ and $b$ is to ``rename'' the variables used internally in the definition of our number.

\subsection{Addition}
Addition can be obtained immediately by noting that the body $sz$ of our definition of the number {\sf 1}, for example, can be interpreted as the application of the function $s$ on $z$. If we want to add say {\sf 2} and {\sf 3}, we just apply the successor function two times to {\sf 3}. 

Let us try the following in order to compute {\sf 2}+{\sf 3}:
$${\sf 2}{\sf S}{\sf 3}\equiv (\lambda sz.s(sz))){\sf S}{\sf 3}\rightarrow {\sf S}({\sf S}{\sf 3})\rightarrow \cdots \rightarrow {\sf 5}$$

In general $m$ plus $n$ can be computed by the expression $m{\sf S}n$.

\subsection{Multiplication}
\begin{figure}[htb]
\centerline{\includegraphics[width=13cm]{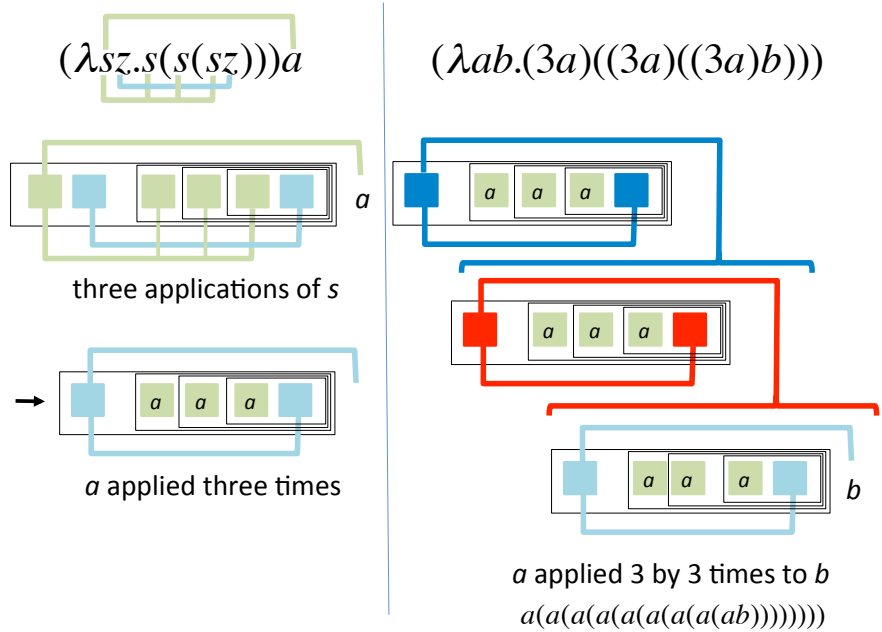}}
\caption{Left: The number {\sf 3} applied to an argument $a$ produces a new function. Right: The plumbing of the function {\sf 3} applied to {\sf 3}a, 
and the result to $b$.\label{mult1}}
\end{figure}

The multiplication of two numbers $x$ and $y$ can be computed using the following function:
$$(\lambda xya.x(ya))$$ The product of {\sf 3} by {\sf 3} is then
$$(\lambda xya.x(ya)){\sf 3}{\sf 3}$$ which reduces to $$(\lambda a.{\sf 3}({\sf 3}a))$$
Using the definition of the number {\sf 3}, we further reduce the above expression to
$$
(\lambda a.(\lambda sb.s(s(sb)))({\sf 3}a)) \rightarrow (\lambda ab. ({\sf 3}a)(({\sf 3}a)(({\sf 3}a)b)))
$$
In order to understand why this function really computes the product of 3 by 3, let us look at some diagrams. The first application 
$({\sf 3}a)$ is computed on the left of  Fig.~\ref{mult1} . Notice that the application of ${\sf 3}$ to $a$ has the effect of producing a new
function which applies $a$ three times to the function's argument.

Now, applying the function ${\sf 3}$ to the result of $({\sf 3}a)$ produces three copies of the function obtained in Fig.~\ref{mult1} , concatenated as shown on the right in Fig.~\ref{mult1} (where the result has been applied to $b$  for clarity). Notice that we have a ``tower'' of three times the same function, each one absorbing the lower one as argument for the application of the function $a$ three times, for a total of nine applications.


\section{Conditionals}
We introduce the following two functions which we call the values ``true''
$${\sf T}\equiv \lambda xy.x$$ and ``false''
$${\sf F}\equiv \lambda xy.y$$
The first function takes two arguments and returns the first one. The second function returns the second of two arguments.

\subsection{Logical operations}
It is now possible to define logical operations using this representation of the truth values.

The {\sf AND} function of two arguments can be defined as
$${\sf \wedge} \equiv \lambda xy.xy {\sf F}$$ 
This definition works because given that $x$ is true, the truth value of the AND operation depends on the truth value of $y$. If $x$ is false (and selects thus the second argument in $xy{\sf F}$) the complete AND is false, regardless of the value of $y$.

The {\sf OR} function of two arguments can be defined as 
$${\sf \vee} \equiv \lambda xy.x {\sf T} y$$
Here, if $x$ is true, the OR is true. If $x$ is false, it picks the second argument $y$ and the value of the OR function depends now on the value of $y$.

Negation of one argument can be defined as 
$$\neg \equiv \lambda x.x {\sf FT}$$

For example, the negation function applied to ``true'' is 
$$\neg{\sf T}\equiv (\lambda x.x {\sf F}{\sf T}){\sf T}$$ which reduces to $${\sf TFT}\equiv (\lambda cd.c){\sf F}{\sf T}\rightarrow {\sf F}$$ that is, the truth value ``false''.

Armed with this three logic functions we can encode any other logic function and reproduce any given circuit without feedback (we look at feedback when we deal with recursion).

\subsection{A conditional test}

It is very convenient in a programming language to have a function which is true if a number is zero and false otherwise. The following function {\sf Z} fulfills this role:
$${\sf Z}\equiv \lambda x.x {\sf F} \neg {\sf F}$$ To understand how this function works, remember that 
$${\sf 0} fa\equiv (\lambda sz.z) fa=a$$ that is, the function $f$ applied zero times to the argument $a$ yields $a$. On the other hand, $F$ applied to any argument yields the identity function
$${\sf F}a\equiv (\lambda xy.y)a \rightarrow \lambda y.y\equiv {\sf I}$$
We can now test if the function {\sf Z} works correctly. The function applied to zero yields
$${\sf Z0} \equiv (\lambda x.x {\sf F} \neg {\sf F}){\sf 0} \rightarrow {\sf 0F}\neg{\sf F} \rightarrow \neg{\sf F} \rightarrow {\sf T}$$ because {\sf F} applied {\sf 0} times to $\neg$ yields $\neg$. The function {\sf Z} applied to any other number {\sf N} yields
$${\sf ZN}\equiv (\lambda x.x{\sf F}\neg{\sf F}){\sf N} \rightarrow {\sf NF}\neg{\sf F} $$ 
The function {\sf F} is then applied {\sf N} times to $\neg$. But {\sf F} applied to anything is the identity (as shown before), so that the above expression reduces, for any number {\sf N} greater than zero, to $${\sf IF} \rightarrow {\sf F}$$

\subsection{The predecessor function}
We can now define the predecessor function combining some of the functions introduced above. When looking for the predecessor of $n$, the general strategy will be to create a pair $(n,n-1)$ and then pick the second element of the pair as the result. 

A pair $(a,b)$ can be represented in $\lambda$-calculus using the function
$$(\lambda z.zab)$$ We can extract the first element of the pair from the expression applying this function to {\sf T}
$$(\lambda z.zab){\sf T}\rightarrow {\sf T} ab\rightarrow a,$$ and the second applying the function to {\sf F}
$$(\lambda z.zab){\sf F} \rightarrow {\sf F} ab\rightarrow b.$$ 
The following function generates from the pair $(n,n-1)$ (which is the argument $p$ in the function) the pair $(n+1,n)$:
$$\Phi\equiv (\lambda pz.z({\sf S}(p{\sf T}))(p{\sf T}))$$
The subexpression $p{\sf T}$ extracts the first element from the pair $p$. A new pair is formed using this element, which is incremented for the first position of the new pair and just copied for the second position of the new pair.

The predecessor of a number $n$ is obtained by applying $n$ times the function $\Phi$ to the pair $(\lambda.z{\sf 00})$ and then selecting the second member of the new pair:
$${\sf P} \equiv (\lambda n.(n\Phi(\lambda z.z{\sf 00})){\sf F})$$
Notice that using this approach the predecessor of zero is zero. This property is useful for the definition of other functions.

\subsection{Equality and inequalities}
With the predecessor function as the building block, we can now define a function which tests if a number $x$ is greater than or equal to a number $y$:
$${\sf G}\equiv (\lambda xy.{\sf Z} (x{\sf P}y))$$
If the predecessor function applied $x$ times to $y$ yields zero, then it is true that $x\geq y$. 

If $x\geq y$ and $y\geq x$, then $x=y$. This leads to the following definition of the function {\sf E} which tests if  two numbers are equal:
$${\sf E}\equiv(\lambda xy.\wedge({\sf Z}(x{\sf P}y))({\sf Z}(y{\sf P}x)))$$
In a similar manner we can define functions to test whether $x>y$, $x<y$ or $x\neq y$.

\section{Recursion}
Recursive functions can be defined in the $\lambda$-calculus using a function which calls a function $y$ and then regenerates itself. This can be better understood by considering the following function {\sf Y}:
$${\sf Y}\equiv (\lambda y.(\lambda x.y(xx))(\lambda x.y(xx)))$$
This function applied to a function {\sf R} yields:
$${\sf YR}\equiv (\lambda x.{\sf R}(xx))(\lambda x.{\sf R}(xx))$$ which further reduced yields
$${\sf R}((\lambda x.{\sf R}(xx))(\lambda x.{\sf R}(xx))$$ 
but this means that ${\sf YR}\rightarrow{\sf R}({\sf YR})$, that is, the function {\sf R} is evaluated using the recursive call {\sf YR} as the first argument.

An infinite loop, for example, can be programmed as {\sf YI}, since this reduces to {\sf I}({\sf YI}), then to {\sf YI} and so ad infinitum.

A more useful function is one which adds the first $n$ natural numbers. We can use a recursive definition, since $\sum^{n}_{i=0}i=n + \sum^{n-1}_{i=0}i$. Let us use the following definition for {\sf R}:
$$R\equiv (\lambda rn.{\sf Z} n{\sf 0}(n{\sf S}(r({\sf P}n))))$$ This definition tells us that the number $n$ is tested: if it is zero the result of the sum is zero. In $n$ is not zero, then the successor function is applied $n$ times to the recursive call (the argument $r$) of the function applied to the predecessor of $n$.

How do we know that $r$ in the expression above is the recursive call to {\sf R}, since functions in $\lambda$-calculus do not have names? We do not know and that is precisely why we have to use the recursion operator {\sf Y}. Assume for example that we want to add the numbers  from {\sf 0} to {\sf 3}. The necessary operations are performed by the call:
$${\sf YR3} \rightarrow  {\sf R}({\sf YR}){\sf 3} \rightarrow {\sf Z30}({\sf 3S}({\sf YR}({\sf P3})))$$ 
Since {\sf 3} is not equal to zero, the evaluation is equivalent to $${\sf 3S}({\sf YR({\sf P3})})$$ that is, the sum of the numbers from {\sf 0} to {\sf 3} is equal to {\sf 3} plus the sum of the numbers from {\sf 0} to the predecessor of {\sf 3} (that is, two). Successive recursive evaluations of {\sf YR} will lead to the correct final result. 

Notice that in the function defined above the recursion will be stopped when the argument becomes {\sf 0}. The final result will be
$$
{\sf 3S}({\sf 2S}({\sf 1S0}))
$$
that is, the number {\sf 6}.


\section{Projects for the reader}
\begin{enumerate} 
\item Define the functions ``less than'' and ``greater than'' of two numerical arguments.
\item Define the positive and negative integers using pairs of natural numbers.
\item Define addition and subtraction of integers.
\item Define the division of positive integers recursively.
\item Define the function $n!=n\cdot (n-1)\cdots1$ recursively.
\item Define the rational numbers as pairs of integers.
\item Define functions for the addition, subtraction, multiplication and division of rationals.
\item Define a data structure to represent a list of numbers.
\item Define a function which extracts the first element from a list.
\item Define a recursive function which counts the number of elements in a list.
\item Can you simulate a Turing machine using $\lambda$-calculus?
\end{enumerate}

 \end{document}